\documentclass{jps-cp}
\usepackage{txfonts} 
\usepackage{amsmath}

\usepackage{ulem}

\usepackage{color}

\def\RE{{\rm Re}} 
\def\IM{{\rm Im}}
\def\Tr{\mathop{\rm Tr}\nolimits}
\def\vecsigma{\boldsymbol{\sigma}}

\def\un{\mathbf{1}} 

\def\pv{{\bf{p}}}

\def\kv{{\bf{k}}}

\def\Sv{{\bf S}}
\def\Vv{{\bf V}}

\def\xu{\hat{\bf x}}
\def\yu{\hat{\bf y}}
\def\zu{\hat{\bf z}}

\def\kt{     { \bf{k}_{r \rm T}  }    }

\def\krt{  \kv_{r\rm T} }                                
\def\krtkrt{\kv^2_{r\rm T}}                     
                           
\def\pt{     { \bf{p}_{r\rm T}  }    }
\def\ptpt{\pv^2_{r \rm T}}

\def\T{_{\rm{T}}}

\def\bt{b_{\rm{T}}}
\def\bl{b_{\rm{L}}}

\def\glgt{|G_{\rm L}/G_{\rm T}|}

\def\gl{G_{\rm L}}
\def\gt{G_{\rm T}}

\title{The String+${}^3P_0$ Model of Hadronization}

\author{Xavier \textsc{Artru}$^{1,*}$, Albi \textsc{Kerbizi}$^{2}$}

\inst{$^{1}$ Universit\'e de Lyon, Institut de Physique des deux Infinis (IP2I Lyon), Universit\'e Lyon 1 and CNRS, 4 rue Enrico Fermi, F-69622 Villeurbanne, France\\
$^{2}$ INFN Sezione di Trieste and Dipartimento di Fisica, Universit\`a degli Studi di Trieste, \\
Via Valerio 2, 34127 Trieste, Italy\\
$^{*}$ Speaker.}

\email{x.artru@ipnl.in2p3.fr, albi.kerbizi@ts.infn.it}

\recdate{January 14, 2022}

\abst{We present the main theoretical aspects of the quantum mechanical string+${}^3P_0$ model of polarized quark fragmentation for pseudoscalar and vector meson production. The model is an extension of the Lund string fragmentation model that systematically includes the quark spin degree of freedom assuming the ${}^3P_0$ mechanism of quark pair production at string breaking. The recently achieved introduction of vector mesons accounts for the spin correlations between the latter and the quark of the fragmentation chain. The string+${}^3P_0$ model reproduces transverse spin effects such as the Collins effect and the dihadron asymmetry, and longitudinal spin effects such as the jet handedness.}

\kword{string model, hadronization, vector mesons, polarization, Collins effect, 3P0}

\begin{document}
\maketitle

\section{Introduction}
Different models are used for the simulation of quark fragmentation, the most successful one being
the Lund string model (LSM) \cite{Lund-model}, where the hadronization is the breaking of a string in smaller string pieces which represent the hadrons. However, most of these models do not account for the quark polarization effects, like the Collins effect \cite{Collins-FF}, which are essential for the quark polarimetry, in particular for the measurement of the transversity distribution in nucleons via semi-inclusive deep inelastic scattering.
A sound model for polarized quark fragmentation is the quantum mechanical string+${}^3P_0$ model. It extends the LSM by systematically including the quark spin degree of freedom assuming that the quark-antiquark pairs produced at the string breakings are in a relative ${}^3P_0$ state. The model is formulated at the amplitude level thus preserving quantum mechanical properties such as positivity and entanglement. It reproduces transverse spin effects such as the Collins effect and the dihadron asymmetry \cite{Bianconi-2h}, and longitudinal spin effects such as the jet handedness \cite{Nachtman-handedness,Efremov-handedness}.
It started with a toy version  \cite{Artru-09} and was developed in Refs \cite{Artru-11,Artru-13}. 
Then it has been implemented in stand alone Monte Carlo programs in two versions, M18 \cite{Kerbizi-2018} and M19 \cite{Kerbizi-2019}, which give similar results.
M19 has also been interfaced to the Pythia 8 hadronization for the simulation of the polarized SIDIS process \cite{StringSpinner}.

In this article we review the string+${}^3P_0$ model and present the main theoretical bases and phenomenological consequences of the recent extension of M19 which includes the vector mesons (VM), calculates their polarizations and  generates their non-isotropic decays \cite{Kerbizi-2021} (version M20).
This new version has also been implemented in a stand alone MC program which allows to study in detail quantitative model predictions on transverse spin asymmetries and to perform comparisons with experimental data. The recipe for the MC implementation is presented in Sec. \ref{sec:recipe}.
The description of the actual implementation as well as the results obtained from simulations with M20 are covered in an other article of these proceedings \cite{these-proceedings}.

\section{The classical string+${}^3P_0$ model}\label{sec:recall}
\begin{figure}[tbh]
\centering
\hspace{-1em}
\begin{minipage}[b]{0.95\textwidth}
\includegraphics[width=1.0\textwidth]{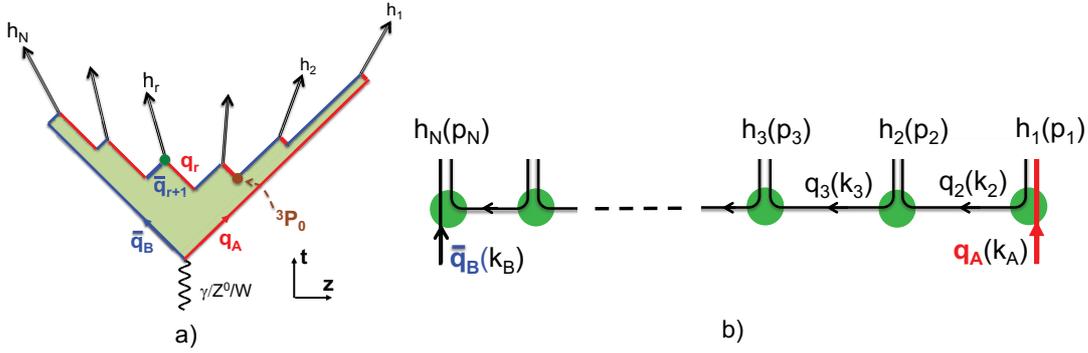}
\end{minipage}\hspace{1em}
\caption{The space-time picture of string fragmentation model (a) and the multiperipheral model of hadronization with quark exchanges (b).}
\label{fig:Models}
\end{figure}

In the string fragmentation model the hadronization of a massive quark-antiquark pair $q_A\bar{q}_B\rightarrow h_1,h_2,\dots h_N$ starts with the creation of a relativistic string stretched between $q_A$ and $\bar{q}_B$.
In the c.m. frame the direction of $q_A$ defines the $\zu$ axis or {\it string axis} (in SIDIS $\bar{q}_B$ represents the target remnant). 
This string undergoes multiple breaking with creation of quark-antiquark pairs, as shown in Fig. \ref{fig:Models}a.
The string piece attached to a quark $q_r$ and an antiquark $\bar{q}_{r+1}$ produced in two adjacent string breaking is an emitted stable or resonant hadron $h_r$, where $r=1,\dots,N$ is the hadron rank (the rank one hadron contains the quark $q_A\equiv q_1$).

The string+${}^3P_0$ model is shown in Fig. \ref{fig:vm_effects}a for pseudoscalar meson (PS) production. It assumes that a created quark-antiquark pair $q_r\bar{q}_r$ is initially in the ${}^3P_0$ state.
The pair has thus unit spin $\textbf{S}=\textbf{s}_q+\textbf{s}_{\bar{q}}$ and unit relative orbital angular momentum $\textbf{L}$ such that the total angular momentum $\textbf{J}=\textbf{L}+\textbf{S}$ vanishes. $\textbf{s}_q$ and $\textbf{s}_{\bar{q}}$ are shown by the small curved arrows in Fig. \ref{fig:vm_effects}a, whereas $L_2,L_3,\dots$ indicate the orbital angular momenta.
The transverse momentum of $h_r$ with respect to the string axis is given by $\textbf{p}_{r\rm T}=\textbf{k}_{r\rm T}- \textbf{k}_{r+1\,T}$, $\textbf{k}_{r\rm T}$ and $-\textbf{k}_{r+1 \rm T}$ being the transverse momenta of the quark and antiquark respectively and shown in Fig. \ref{fig:vm_effects}a by the straight arrows.

Assuming $q_A$ polarized along the $\yu$ axis, odd and even rank hadrons have on the average opposite transverse momenta along the $\xu$ axis, resulting in Collins effects of alternate signs. If $q_A$ is a $u$ quark, negative and positive Collins effects are obtained for $\pi^+$ and $\pi^-$ respectively, in agreement with experiment.

\section{Quantum mechanical formulation}\label{sec:TSA}\label{sec:quantum mechanical}
\subsection{The multiperipheral model}
The quantum mechanical formulation of the string+${}^3P_0$ model is achieved by casting the string decay process in Fig. \ref{fig:Models}a in a multiperipheral diagram with quark exchanges \cite{Artru-11}, shown in Fig. \ref{fig:Models}b. 
The quark spin states are described by Pauli spinors obtained from the reduction of the corresponding Dirac spinors in the $\alpha_z = v_z = \pm1$ subspace. $v_z$ = -1 (+1) for a quark entering (leaving) a hadron, according to Fig. \ref{fig:Models}a. The diagram in Fig. \ref{fig:Models}b can be viewed as the recursive repetition of the elementary splitting $q_r\rightarrow h_r+q_{r+1}$ which is described in flavor, momentum and spin space by the splitting amplitude $T_{q_{r+1},h_r,q_r}(k_{r+1},p_r,k_r)= \Delta_{q_{r+1}}(k_{r+1}) \times \Gamma_{q_{r+1},h_r,q_r}(k_{r+1},p_r,k_r)$ is composed by the quark propagator $\Delta_{q_{r+1}}$ and the quark meson coupling $\Gamma_{q_{r+1},h_r,q_r}$, which are $2\times 2$ matrices in the quark spin space. The variables $k_r$, $p_r$ and $k_{r+1}=k_r-p_r$ indicate the four momenta of $q_r$, $h_r$ and $q_{r+1}$ respectively. The spin information of $q_r$ is encoded in the $2\times 2$ density matrix $\rho(q_r)=(\un+\vecsigma\cdot \textbf{S}_{q_r})/2$, where $\vecsigma$ is the vector of Pauli matrices and $\textbf{S}_{q_r}$ is the polarization vector of $q_r$.

The multiperipheral amplitude must respect the \textit{Left-Right} (or \textit{quark Line Reversal}) symmetry, namely the diagram in Fig. \ref{fig:Models}b can be thought to evolve from $q_A$ towards $\bar{q}_B$ via quark exchanges or from $\bar{q}_B$ towards $q_A$ via antiquark exchanges with the same probability. This symmetry is respected by the LSM \cite{Lund-model} and is also a non trivial constraint on the spin-dependent parts of the splitting amplitude. The latter is obtained by the following ingredients.

The spin-dependent part of the propagator $\Delta_{q_r}$ is obtained from the ${}^3P_0$ wave function of the $q_{r}\bar{q}_{r}$ pair. It correlates the spin $\Sv$ of $q_r$ going toward $h_r$ to that $\bar\Sv$ of $\bar q_r$ going toward $h_{r-1}$ or $\Sv'\equiv-\bar\Sv$ of $q_r$ leaving $h_{r-1}$. In the $k_r^0=0$ frame this wave function is proportional to $\langle q_r(k_r,\Sv),\bar q_r(-k_r,\bar\Sv)|{}^3P_0\rangle = \chi^{\dagger}(\Sv)\,\sigma_z\vecsigma\cdot \kv_r\,\chi(\Sv')$, where $\chi(\textbf{S})$ is the Pauli spinor of polarization $\textbf{S}$.

During tunneling the relative momentum of the pair $\kv_{r}$ has an imaginary component along $\zu$ given by $k^{z}_{r}\sim i\,(m_{q_r}^2+\krtkrt)^{1/2}$, where $m_{q_{r}}$ is the quark mass. Replacing $k^{z}_{r}$ by a fixed complex mass parameter $\mu$ the spin part of the propagator becomes $\mu+\sigma_z\vecsigma\cdot \krt$ which is the analogue of the spin part of the Feynman propagator $m_q+\gamma\cdot k$.~ $\RE\mu$ and $\IM\mu$ are free parameters of the string+${}^3P_0$ model. $\IM\mu\ne0$ leads to Collins effect and dihadron asymmetry, while $\RE\mu\times\IM\mu\ne0$ leads to jet handedness.

The coupling matrix $\Gamma_{h_r}$ implements the correlations between the polarizations of $q_r$ and $\bar{q}_{r+1}$ constituting $h_r$. For a PS meson ($\pi$, $K$, $\eta$, $\eta'$), the pair $q_r\bar{q}_{r+1}$ is in the spin singlet state. This correlation is implemented by $\Gamma^{PS}_{h_r}=\sigma_z$, which is the analogue of the $\gamma_5$ matrix in Dirac space. For a VM ($\rho$, $K^*$, $\omega$, $\phi$), the coupling depends on the vector amplitude $\textbf{V}$ of the meson. The most simple form which respects LR symmetry and parity is $\Gamma_{h_r}^{VM}=\gl V_z^*+\gt\textbf{V}_{\rm T}^* \cdot \vecsigma\sigma_z$. The couplings constants $\gl$ and $\gt$ are free parameters and can be complex valued. However the relevant free parameters are $2|\gt|^2+|\gl|^2$ related to the abundance of VMs, $\glgt^2$ responsible for the alignment (and Collins effect) of VMs and $\arg{\glgt}$ responsible for the oblique polarization of VMs.

All considered, the final expression for the splitting amplitude respecting the LR symmetry for the splitting $q_r\rightarrow h_r+q_{r+1}$ (more shortly rewritten $q\to h+q'$) is \cite{Kerbizi-2021}
\begin{eqnarray}\label{eq:T}
T_{q',h,q} = N\,
\left[
(1-Z)^a e^{-\bl\frac{m_{h}^2+\pv\T^2}{Z}-\bt{\kv'}^2\T}
\right]^{\frac{1}{2}} 
 (\mu+\sigma_z\vecsigma\cdot\kv'\T)\times \begin{cases}
\sigma_z, & \text{$h$=PS}. \\
\gl V_z^*+\gt\Vv\T^* \cdot \vecsigma\,\sigma_z, & \text{$h$=VM}.
\end{cases}
\end{eqnarray}
%
%
%
$Z=p^+/k^+$ is the forward lightcone momentum fraction and $N$ is a normalization factor. The quantity in square bracket is the splitting function of the spin-blind LSM. The free parameters $a$, $\bl$ and $\bt$ are the same as in the LSM and account for the quark actions, the string area law and the tunneling of $q\bar{q}$ pairs at string breaking points respectively \cite{Artru-11}. For VM production, Eq. (\ref{eq:T}) must be multiplied also by a spectral function to describe the distribution of the invariant mass $m_{h}$ of the resonance.

\subsection{Effects related to the vector meson polarization}
\begin{figure}[tbh]
\centering
\begin{minipage}[b]{1.0\textwidth}
\includegraphics[width=1.0\textwidth]{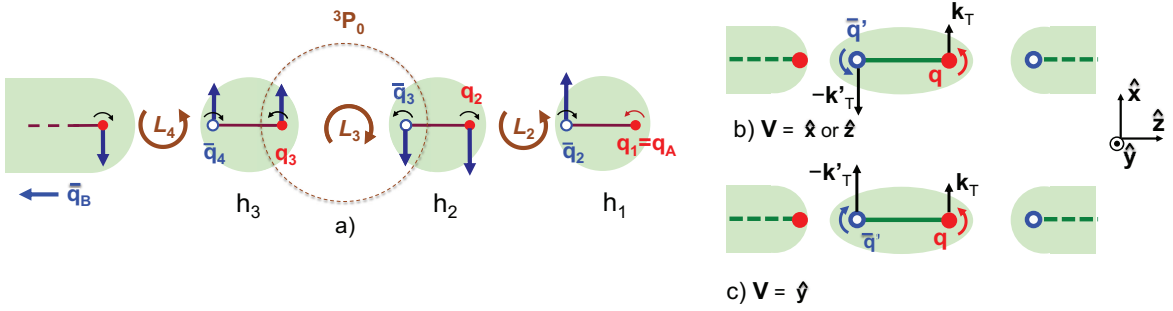}
\end{minipage}
\caption{The classical string+${}^3P_0$ model applied to PS meson production (a). The string+${}^3P_0$ model with production of a VM with linear polarization along $\xu$ or $\zu$ axis (b) and with linear polarization along $\yu$ axis (c).}
\label{fig:vm_effects}
\end{figure}

The dependence of the Collins effect for a VM $h=q\bar{q}'$ on the polarization vector $\textbf{V}$ is shown in Fig. \ref{fig:vm_effects}b for $\textbf{V}=\xu$ or $\zu$, and in Fig. \ref{fig:vm_effects}c for $\textbf{V}=\yu$. In both figures the spin of $q$ is along $\yu$. If the VM is polarized along $\xu$ or $\zu$, the spin projections $S_y(\bar{q}')$ and $S_y(q)$ have the same sign and the VM has opposite Collins effect compared to PS case. If the VM is polarized along $\yu$ axis, $S_y(\bar{q}')$ and $S_y(q)$ have opposite signs, so that the $1^{\rm rst}$-rank VM behaves as a PS meson (compare with Fig. \ref{fig:vm_effects}a).
All in all, applied to rank one, VMs have opposite Collins effect compared to PS mesons in agreement with \cite{Czyzewski}. Also, for larger ranks the VM have smaller average transverse momenta squared compared to PS mesons.

Fig. \ref{fig:flow}a shows instead the effect of the decay of a VM with oblique polarization for a $\rho$ meson decaying in a pair of pions (the ellipse represents the oblique polarization). When the decay products are ordered according to their fraction of fragmenting quark energy, the oblique polarization is a source of dihadron asymmetry \cite{Kerbizi-2021}. It is also a source of Collins effect for the final observed hadrons \cite{these-proceedings}.

\section{Recipe for the Monte Carlo implementation}\label{sec:recipe}
The Monte Carlo implementation of the polarized quark fragmentation chain bears on the recursive repetition of the elementary splitting $q_r\rightarrow h_r + q_{r+1}$. Assuming that the flavor, the momentum and the density matrix $\rho(q_r)$ of $q_r$ are known, the generation of the next elementary splitting is achieved by the following steps:
\begin{itemize}
\item[i.] Generate the type and the four-momentum of $h_r$ by using the probability density $dP_{q_{r+1},h_r,q_r}=F_{q_{r+1},h_r,q_r}(Z,\pt;\kt,\textbf{S}_{q_r})\,dZZ^{-1}\,d^2\pt$ where the splitting function is obtained as
\begin{equation}
    F_{q_{r+1},h_r,q_r}=\begin{cases}
    \Tr T_{q_{r+1},h_r,q_r}\,\rho(q_r)\,T^{\dagger}_{q_{r+1},h_r,q_r}, & \text{$h_r$ = PS}.\\
    \sum_{\alpha}^{}\,\Tr\,T^{\alpha}_{q_{r+1},h_r,q_r}\,\rho(q_r)\,T^{\alpha\,\dagger}_{q_{r+1},h_r,q_r}, & \text{$h_r$=VM of spin state $|\alpha\rangle$}.
  \end{cases} 
\end{equation}
The trace operation is performed over the quark spin indices. In the VM case the summation is performed over the vector meson polarization states $\alpha=\xu,\yu,\zu$, after writing the splitting amplitude as $T_{q_{r+1},h_r,q_r}=T_{q_{r+1},h_r,q_r}^{\alpha}V^{\alpha}$.
\item[ii.] For a PS meson or not analyzed VM $h_r$ go to step vi. For an analyzed VM continue with step iii.
\item[iii.] Calculate the density matrix of the VM as $\rho_{\alpha\beta}(h_r) = \Tr\,T^{\alpha}_{q_{r+1},h_r,q_r}\,\rho(q_r)\,T^{\beta\,\dagger}_{q_{r+1},h_r,q_r}/F_{q_{r+1},h_r,q_r}^{-1}$.
\item[iv.] Simulate the anisotropic decay of the VM by generating in the meson rest frame the relative momentum $\hat{\textbf{r}}$ of the decay products according to the angular distribution $dN(\hat{\textbf{r}})/d\Omega = \langle \hat{\textbf{r}}|\mathcal{M}|\alpha\rangle\,\rho_{\alpha\beta}(h_r)\,\langle \beta|\mathcal{M}^{\dagger}|\hat{\textbf{r}}\rangle$, where $\langle \hat{\textbf{r}}|\mathcal{M}|\alpha\rangle$ is the decay amplitude. 
Go to the string rest frame by applying to the decay products 
 successively the boosts of velocities $\pt/(m_{h_r}+\ptpt)^{1/2})$ and $(p^z_r/p^0_r)\,\zu$.
This satisfies the left-right symmetry.
\item[v.] Calculate the decay matrix $\mathcal{D}_{\beta\alpha}=\langle \beta|\mathcal{M}^{\dagger}|\hat{\textbf{r}}\rangle\,\langle \hat{\textbf{r}}|\mathcal{M}|\alpha\rangle$.
\item[vi.] Calculate the spin density matrix of $q_{r+1}$ by the rule \begin{equation}
\rho(q_{r+1})=\begin{cases}
T_{q_{r+1},h_r,q_r}\,\rho(q_r)\,T^\dagger_{q_{r+1},h_r,q_r}, & \text{$h_r$=PS}. \\
\sum_{\alpha}^{}\,T^{\alpha}_{q_{r+1},h_r,q_r}\,\rho(q_r)\,T^{\alpha\,\dagger}_{q_{r+1},h_r,q_r}, & \text{$h_r$=not analyzed VM}. \\
\mathcal{D}_{\beta\alpha}\,\Tr\,T^{\alpha}_{q_{r+1},h_r,q_r}\,\rho(q_r)\,T^{\beta\,\dagger}_{q_{r+1},h_r,q_r}, & \text{$h_r$=analyzed VM}.
\end{cases}
\end{equation}
\end{itemize}

\begin{figure}[tbh]
\centering
\begin{minipage}[b]{0.6\textwidth}
\includegraphics[width=1.0\textwidth]{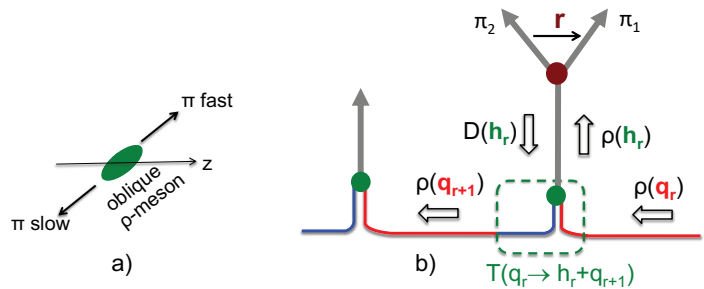}
\end{minipage}
\caption{Decay of a $\rho$ meson with oblique polarization (a). Path of the spin information (shown by the empty arrows) from the fragmenting quark $q_r$ to the decay event of $h_r$, back to the splitting event and finally to the leftover quark $q_{r+1}$ (b).}
\label{fig:flow}
\end{figure}

The steps i - vi are represented in Fig. \ref{fig:flow}b where the arrows indicate the path of the spin information from the fragmenting quark $q_r$ to the quark $q_{r+1}$ via the production and decay of the VM $h_r$. The application of steps iii to v is necessary to account for the correlations between the polarization of $q_{r+1}$ and the direction of the decay products of $h_r$. It is not allowed to treat separately the decay of $h_r$ and the fragmentation of $q_{r+1}$. The decay matrix $\mathcal{D}_{\beta\alpha}$ must be employed to carry the information about the direction of $\hat{\textbf{r}}$ from the decay event to the splitting event, in violation of classical causality \cite{Collins-corr,Knowles-corr}.

\section{Conclusions}
The Lund string fragmentation model has been extended by systematically including quark spin effects for PS and VM productions. Quark spin effects are introduced by assuming the ${}^3P_0$ mechanism of quark pair production at string breaking, which is parametrized by a complex mass parameter responsible for transverse (Collins effect, dihadron asymmetry) and longitudinal (jet handedness) spin effects. The introduction of spin effects for VMs require the introduction of three new free parameters responsible for the abundance of VMs, the alignment (and Collins effect) of the VMs and their oblique polarizations. Vector mesons are shown to have opposite Collins effect with respect to PS mesons. More detailed results have been obtained from the stand alone MC implementation of the model and are shown in Ref. \cite{these-proceedings}.

\end{document}